\documentclass[11pt,a4paper]{article}
\usepackage{amsmath,amsfonts,amssymb,amscd}
\catcode`\@=11
\@addtoreset{equation}{section}

\catcode`\@=12

\newfont{\gotico}{eufm10 scaled\magstephalf}
\newfont{\qvd}{msam10 scaled\magstephalf}

\def\de#1/de#2{\frac{\partial {#1}}{\partial {#2}}}
\def\De#1/de#2{\dfrac{\partial {#1}}{\partial {#2}}}

\begin{document}
\vskip-2cm
\title{Dirac fields in $f(R)$-gravity with torsion}
\author{Luca Fabbri\footnote{E-mail: luca.fabbri@bo.infn.it}\\
\footnotesize{INFN \& Dipartimento di Fisica, Universit\`{a} di Bologna}\\
\footnotesize{Via Irnerio 46, - 40126 Bologna (Italia) and}\\
\footnotesize{DIPTEM Sez. Metodi e Modelli Matematici, Universit\`{a} di Genova}\\
\footnotesize{Piazzale Kennedy, Pad. D - 16129 Genova (Italia)}
\and Stefano Vignolo\footnote{E-mail: vignolo@diptem.unige.it}\\
\footnotesize{DIPTEM Sez. Metodi e Modelli Matematici, Universit\`{a} di Genova}\\
\footnotesize{Piazzale Kennedy, Pad. D - 16129 Genova (Italia)}}
\date{}
\maketitle
\begin{abstract}
\noindent
We study $f(R)$-gravity with torsion in presence of Dirac massive fields. Using the Bianchi identities, we formulate the conservation laws of the theory and we check the consistency with the matter field equations. Further, we decompose the field equations in torsionless and torsional terms: we show that the non-linearity of the gravitational Lagrangian reduces to the presence of a scalar field that depends on the spinor field; this additional scalar field gives rise to an effective stress-energy tensor and plays the role of a scale factor modifying the normalization of Dirac fields. Problems for fermions regarding the positivity of energy and the particle-antiparticle duality are discussed.
\end{abstract}
\section{Introduction}
In the last thirty years, several extensions of General Relativity (GR) have been formulated in order to solve some problems left open by Einstein's theory at ultra-violet and infra-red scales; among all the developed extended theories of gravity, one of the simplest approaches is certainly given by the so-called $f(R)$-theory: it consists in relaxing the hypothesis that the gravitational Lagrangian be a linear function of the Ricci scalar $R$ allowing a dependence on $R$ more general than that of the Einstein-Hilbert action. This approach has acquired great interest in cosmology and astrophysics where $f(R)$-gravity turned out to be useful in addressing cosmological and astrophysical puzzles such as dark energy and dark matter; for example, it leads to alternative explanations of the accelerated behaviour of the today universe as well as the missing matter at astrophysical scales without postulating the existence of forms of energy and matter unknown so far.
\\
Further reasons to investigate extended theories of gravity come from the Field Theory; as it is well-known, in dealing with elementary particles the spin is as fundamental as the momentum. In order to take spin into account, first Cartan introduced torsion in the geometrical background, and then Sciama and Kibble took it to the framework of Einstein gravity following the scheme for which the energy is the source of curvature in the same way in which the spin is the source of torsion; in the resulting Einstein-Cartan-Sciama-Kibble (ECSK) theory, the gravitational fields are given by a metric tensor and a metric compatible connection with torsion, whose dynamics is described by a gravitational Lagrangian linear in the scalar curvature. In this scenario, $f(R)$-gravity with torsion has been developed as one of the simplest extensions of the ECSK theory, the one relying again on the idea of replacing the Einstein-Hilbert Lagrangian with a non-linear function. An important consequence of this fact is the presence of torsion even without spin as long as the stress-energy trace is not constant \cite{CCSV1,CCSV2,CCSV3,CV4}. Nonetheless, the matter field that best exploits the coupling between torsion and spin, and therefore the most general to study, is the spinor field, the simplest of which is the spin-$\frac{1}{2}$ spinor field, already studied in the special case of masslessness by Rubilar \cite{Rubilar}.
\\
In this paper, we consider $f(R)$-gravity with torsion coupled to the Dirac spinor field in its most general massive case; our approach will face the problem of the consistency of matter field equations with the conservation laws. In fact, the ECSK-like field equations of the theory involve the curvature and the torsion tensors which appear also in the well-known Bianchi identities; the latter are geometrical and they hold independently on the ECSK-like field equations through which they are converted into suitable conservation laws; these conservation laws must be satisfied once matter field equations are eventually given. A point that has to be stressed is that the obtained conservation laws reduce to the ones of the Einstein theory when the spin vanishes (for instance, the particles of a perfect fluid without spin follow the Levi--Civita geodesics, even if the geodesic structure of spacetime differs from the Levi--Civita one). However, in presence of spin all equations are more general than those of the Einstein theory, and for them the check of consistency may not be trivial (for example, for some of the higher-spin supersymmetric models, constraints known as subsidiary conditions need be added in order for such consistency to be obtained); this is a fundamental aspect for the viability of the theory. In particular, in the case of $f(R)$-gravity coupled to the Dirac field, we verify that the Dirac-like field equations are actually consistent with the aforementioned conservation laws by checking that the former automatically imply the validity of the latter; for this model, as far as our knowledge is concerned, this consistency check has never been carried out in the literature.
\\
After checking the consistency of the field equations, we shall further proceed by decomposing all of them into torsionless and torsional terms, showing that the contributions due to the non-linearity of the function $f(R)$ are equivalent to the presence of a suitable scalar field depending on Dirac field alone. This scalar field is seen to modify the effective stress-energy tensor of the theory and it enters all equations as a scale factor modifying the normalization of Dirac field. We shall see that this fact has consequences for the Dirac field itself: indeed, the possibility to have the two opposite signs for the scale factor that multiplies the terms containing the Dirac field implies that it is possible to have both repulsive and attractive spin-torsion coupling within the Dirac-like field equations and to have both the negative and positive energy, respectively for particles and antiparticles, in the Einstein-like field equations; in particular, this last point is important for the problem of the positivity of the energy of particles and antiparticles, because it solves it in the classical domain. On the same footing, this situation leaves open the path to the possibility of having a particle-antiparticle duality lost in gravitational physics.
\section{Geometrical preliminaries}
Throughout the paper we indicate spacetime indices by Latin letters. A metric tensor on the spacetime $M$ is denoted by $g_{ij}$ and a connection by $\Gamma_{ij}^{\;\;\;h}$; metric-compatible connections are those whose covariant derivative applied on the metric tensor vanishes. Any connection $\Gamma_{ij}^{\;\;\;h}$ defines the associated covariant derivative
\begin{equation}
\label{2.1}
\nabla_{\partial_i}\partial_j = \Gamma_{ij}^{\;\;\;h}\,\de /de{x^h}
\end{equation}
as well as the corresponding torsion and Riemann curvature tensors expressed as
\begin{subequations}
\label{2.2}
\begin{equation}
\label{2.2a}
T_{ij}^{\;\;\;h}
=\Gamma_{ij}^{\;\;\;h}-\Gamma_{ji}^{\;\;\;h}
\end{equation}
\begin{equation}
\label{2.2b}
R^{h}_{\;\;kij}
=\partial_i\Gamma_{jk}^{\;\;\;h} - \partial_j\Gamma_{ik}^{\;\;\;h} +
\Gamma_{ip}^{\;\;\;h}\Gamma_{jk}^{\;\;\;p}-\Gamma_{jp}^{\;\;\;h}\Gamma_{ik}^{\;\;\;p}
\end{equation}
\end{subequations}
where contractions $T_{i}= T_{ij}^{\;\;\;j}$, $R_{ij}=R^{h}_{\phantom{h}ihj}$ and $R=R_{ij}g^{ij}$ are called respectively the torsion vector, the Ricci tensor and the Ricci scalar curvature. The torsion and curvature tensors are used to express the commutator of covariant derivatives as 
\begin{equation}
\label{2.4}
[\nabla_{i},\nabla_{j}]V_{k}=-T_{ij}^{\;\;\;h}\nabla_{h}V_{k}-R^{a}_{\;\;kij}V_{a}
\end{equation}
for any generic vector $V_{k}$. By considering the commutators of commutators in cyclic permutation one can obtain the so-called Bianchi identities 
\begin{subequations}
\label{2.5}
\begin{eqnarray}
\label{2.5b}
\nonumber
&\nabla_{c}T_{ij}^{\;\;\;h}+\nabla_{i}T_{jc}^{\;\;\;h}+\nabla_{j}T_{ci}^{\;\;\;h}
-T_{ij}^{\;\;\;a}T_{ca}^{\;\;\;h}-T_{jc}^{\;\;\;a}T_{ia}^{\;\;\;h}-T_{ci}^{\;\;\;a}T_{ja}^{\;\;\;h}-\\
&-R^{h}_{\;\;cij}-R^{h}_{\;\;ijc}-R^{h}_{\;\;jci}=0
\end{eqnarray}
\begin{eqnarray}
\label{2.5a}
\nabla_{c}R^{p}_{\;\;kij}+\nabla_{i}R^{p}_{\;\;kjc}+\nabla_{j}R^{p}_{\;\;kci}
-T_{ij}^{\;\;\;a}R^{p}_{\;\;kca}-T_{jc}^{\;\;\;a}R^{p}_{\;\;kia}-T_{ci}^{\;\;\;a}R^{p}_{\;\;kja}=0
\end{eqnarray}
\end{subequations}
whose usefulness will be clear in the following.
\\
Given a metric tensor $g_{ij}$, every metric $g$-compatible connection can be decomposed in the form \cite{HD,HHKN}
\begin{equation}
\label{2.6}
\Gamma_{ij}^{\;\;\;h}=\tilde{\Gamma}_{ij}^{\;\;\;h}-K_{ij}^{\;\;\;h}
\end{equation}
where 
\begin{equation}
\label{2.7}
K_{ij}^{\;\;\;h}
=\frac{1}{2}\left(-T_{ij}^{\;\;\;h}+T_{j\;\;\;i}^{\;\;h}-T^{h}_{\;\;ij}\right)
\end{equation}
is called the contorsion tensor and $\tilde{\Gamma}_{ij}^{\;\;\;h}$ is the Levi--Civita connection induced by the metric $g_{ij}$; the contorsion tensor satisfies the antisymmetry property given by $K_{i}^{\;\;jh}=-K_{i}^{\;\;hj}$ amounting to the metric-compatibility of every connection of the form \eqref{2.6}. The contorsion has one contraction $K_{i}^{\;\;ij}=K^{j}$, from which we have also the identity $K_{i}=-T_{i}$. With the contorsion we can decompose the covariant derivative of the full connection into covariant derivative of the Levi--Civita connection $\tilde{\nabla}\/$ plus contorsional contributions; this gives the following representation of the Ricci curvature
\begin{equation}
\label{2.8}
R_{ij}=\tilde{R}_{ij} + \tilde{\nabla}_jK_{hi}^{\;\;\;h} - \tilde{\nabla}_hK_{ji}^{\;\;\;h} + K_{ji}^{\;\;\;p}K_{hp}^{\;\;\;h} - K_{hi}^{\;\;\;p}K_{jp}^{\;\;\;h}
\end{equation}
with the Ricci curvature of the Levi--Civita connection $\tilde{R}_{ij}\/$ and contorsional terms.
\\
In the next sections we shall consider $f(R)$-gravity coupled with Dirac fields: as it is well known, the most suitable variables to describe fermion fields interacting with gravity are tetrad fields. Tetrad fields possess Lorentz indices denoted by Greek letters as well as spacetime indices. They are defined by $e^{\mu}=e^{\mu}_{i}\,dx^{i}$ together with their dual $e_{\mu}=e_{\mu}^{i}\,\de /de{x^{i}}$ where $e^{j}_{\mu}e^{\mu}_{i}=\delta^{j}_{i}$ and $e^{j}_{\mu}e^{\nu}_{j}=\delta^{\nu}_{\mu}$ and spin-connections $1$-forms $\omega^\mu_{\;\;\nu} = \omega_{i\;\;\;\nu}^{\;\;\mu}\,dx^i$. Metric compatibility conditions are assumed and they are defined by the requirement that the covariant derivatives of tetrads and Minkowskian metric vanish, respectively implying that 
\begin{equation}
\label{2.9}
\Gamma_{ij}^{\;\;\;h} = \omega_{i\;\;\;\nu}^{\;\;\mu}e_\mu^h\/e^\nu_j + e^{h}_{\mu}\partial_{i}e^{\mu}_{j}
\end{equation}
and the antisymmetry  $\omega_{i}^{\;\;\mu\nu}=-\omega_{i}^{\;\;\nu\mu}$ of the spin connection.
\\
The torsion and curvature tensors induced by the assignment of a spin-connection together with a tetrad field are expressed as 
\begin{subequations}
\label{2.10}
\begin{equation}
\label{2.10a}
T^\mu_{ij}
=\partial_i\/e^\mu_j - \partial_j\/e^\mu_i
+\omega^{\;\;\mu}_{i\;\;\;\lambda}e^\lambda_{j}
-\omega^{\;\;\mu}_{j\;\;\;\lambda}e^\lambda_{i}
\end{equation}
\begin{equation}
\label{2.10b}
R_{ij}^{\;\;\;\;\mu\nu}
=\partial_i\omega_{j}^{\;\;\mu\nu} - \partial_j{\omega_{i}^{\;\;\mu\nu}}
+\omega^{\;\;\mu}_{i\;\;\;\lambda}\omega_{j}^{\;\;\lambda\nu}
-\omega^{\;\;\mu}_{j\;\;\;\lambda}\omega_{i}^{\;\;\lambda\nu}
\end{equation}
\end{subequations}
and their relationships with the world tensors defined in equations \eqref{2.2} are given by the relationships $T_{ij}^{\;\;\;h}:=T^{\;\;\;\alpha}_{ij}e_\alpha^h$ and $R^{h}_{\;\;kij}=R_{ij\;\;\;\;\nu}^{\;\;\;\;\mu}e_\mu^h\/e^\nu_k$ respectively.
\section{$f(R)$-gravity with torsion and conservation laws}
$f(R)$-gravity with torsion can be formulated in the metric-affine approach \cite{CCSV1} or in the tetrad-affine one \cite{CCSV2}. In the first case, the gravitational dynamical fields are represented by the metric $g\/$ and a metric compatible connection $\Gamma\/$ on the spacetime manifold $M\/$. In the second case, the gravitational dynamical fields are given by a tetrad field $e^\mu_i$ and a spin-connection $\omega_{i}^{\;\;\mu\nu}$, still defined on $M$. Field equations are derived variationally through a Lagrangian function of the kind 
\begin{equation}
\label{3.1}
{\cal L}=f(R)-{\cal L}_m
\end{equation}
where $f(R)$ is a real function of the Ricci curvature scalar $R$ written in terms of the metric and connection, or equivalently tetrad and spin-connection, and ${\cal L}_m$ indicates a suitable matter Lagrangian. 
\\
In general the metric-affine formulation is preferred when the connection does not couple with matter, namely when the matter Lagrangian is independent of the connection. In such a circumstance, the field equations of the theory are \cite{CCSV1}
\begin{subequations}
\label{3.2}
\begin{equation}
\label{3.2a}
f'\/(R)R_{ij} - \frac{1}{2}f\/(R)g_{ij}=\Sigma_{ij}
\end{equation}
\begin{equation}
\label{3.2b}
T_{ij}^{\;\;\;h} = -
\frac{1}{2f'\/(R)}\de{f'\/(R)}/de{x^p}\/\left(\delta^p_i\delta^h_j
- \delta^p_j\delta^h_i\right)
\end{equation}
\end{subequations}
where $\Sigma_{ij}:= \frac{1}{\sqrt{|g|}}\frac{\delta{\cal L}_m}{\delta g^{ij}}$ plays the role of the stress-energy tensor. Equations \eqref{3.2a} are by construction symmetric. The symmetrization symbol for the Ricci tensor $R_{ij}$ is omitted because any connection satisfying equations \eqref{2.6}, \eqref{2.7} and \eqref{3.2b} gives rise to a Ricci tensor that is automatically symmetric \cite{CCSV1,CCSV3,CV4}. We also notice that, if $f(R)\not = kR$, we have non-vanishing torsion even if the spin density is zero. 
\\
On the other hand, in case of coupling with spin matter, for instance spinor fields, the tetrad-affine formulation is more suitable. The corresponding field equations have been derived in \cite{CCSV2} (see also \cite{Rubilar}) and turn out to be 
\begin{subequations}
\label{3.3}
\begin{equation}
\label{3.3a}
f'\/(R)R_{\mu\sigma}^{\;\;\;\;\lambda\sigma}e^{i}_\lambda -\frac{1}{2}e^i_{\mu}f\/(R)=\Sigma^i_\mu
\end{equation}
\begin{equation}
\label{3.3b}
T_{ts}^{\;\;\;\alpha}
=\frac{1}{f'(R)}
\left[\frac{1}{2}\left(\de{f'(R)}/de{x^{p}}+S_{p\sigma}^{\;\;\;\sigma}\right)
\left(\delta^{p}_{s}e^{\alpha}_{t}-\delta^{p}_{t}e^{\alpha}_{s}\right)
+S_{ts}^{\;\;\;\alpha}\right]
\end{equation}
\end{subequations}
where $\Sigma^{i}_{\;\mu}:=-\frac{1}{2e}\frac{\partial{\cal L}_m}{\partial e^{\mu}_{i}}$ and $S_{ts}^{\;\;\;\alpha}:=\frac{1}{2e}\frac{\partial{\cal L}_m}{\partial\omega_{i}^{\;\;\mu\nu}}e^{\mu}_{t}e^{\nu}_{s}e^{\alpha}_{i}$ are the stress-energy and spin density tensors of the matter field. From equation \eqref{3.3b} it is seen that, in presence of $\omega$-dependent matter, there are two sources of torsion: the spin density $S_{ts}^{\;\;\;\alpha}\/$ and the non-linearity of the gravitational Lagrangian.
\\
By considering the trace of equations \eqref{3.2a} or \eqref{3.3a} we obtain the relation 
\begin{equation}
\label{3.4}
f'\/(R)R -2f\/(R) = \Sigma
\end{equation}
between the curvature scalar $R\/$ and the trace $\Sigma\/$ of the stress-energy tensor.
\\
From \eqref{3.4} we see that when the trace $\Sigma\/$ is only allowed to be constant (where the constant can assume any value, in particular zero) then also $R$ is forced to be constant; in such a circumstance, the present theory amounts to an Einstein-like (if $S^{\;\;\;\alpha}_{ts}=0$, i.e. $\omega$-independent matter) or an ECSK-like (if $S^{\;\;\;\alpha}_{ts}\not= 0$, i.e. $\omega$-dependent matter) theory with cosmological constant \cite{CCSV1,CCSV2,CCSV3,Rubilar}. The previous statement holds with the exception of the particular case given, when $\Sigma=0\/$, by the relationship $f(R)=k\/R^2\/$, because under these conditions equation \eqref{3.4} is a trivial identity imposing no restrictions on the curvature scalar $R\/$. 
\\
Therefore, from now on we shall systematically assume that $\Sigma\/$ is not constant as well as that $f(R)\not = kR^2$. Moreover, we shall assume that the relation \eqref{3.4} is invertible so that the Ricci curvature is a suitable function of $\Sigma\/$, namely $R=F\/(\Sigma)$. 
\\
With the stated assumptions in mind, defining the world tensors $R^i_{\;j}:=R_{\mu\sigma}^{\;\;\;\;\lambda\sigma}e^{i}_{\lambda}e^\mu_j$, $\Sigma^i_{\;j}:=\Sigma^i_{\mu}e^\mu_j$, $T_{ij}^{\;\;\;h}:=T^{\;\;\;\alpha}_{ij}e_\alpha^h\/$ and $S^{\;\;\;h}_{ij}:=S^{\;\;\;\alpha}_{ij}e_\alpha^h\/$, we can rewrite equations \eqref{3.3} in their equivalent general covariant form 
\begin{subequations}
\label{3.5}
\begin{equation}
\label{3.5a}
R_{ij}
=\frac{1}{f'(F(\Sigma))}
\left[\frac{1}{4}\left(f'(F(\Sigma))F(\Sigma) - \Sigma\right)g_{ij} + \Sigma_{ij}\right]
\end{equation}
\begin{equation}
\label{3.5b}
T_{ij}^{\;\;\;h} = - \frac{1}{2f'\/(F\/(\Sigma))}\left(\de{f'\/(F\/(\Sigma))}/de{x^p} + S_{pq}^{\;\;\;q}\right)\left(\delta^p_i\delta^h_j - \delta^p_j\delta^h_i\right) + \frac{1}{f'\/(F\/(\Sigma))}S^{\;\;\;h}_{ij}
\end{equation}
\end{subequations}
giving the Ricci curvature tensor and torsion tensor in terms of the energy and spin densities. In equations \eqref{3.5a} one has to distinguish the order of the indices since in general the tensors $R_{ij}\/$ and is not symmetric if $S^{\;\;\;h}_{ij}\not=0$ while $\Sigma_{ij}\/$ is not symmetric because it is obtained by variation with respect to the tetrad fields. It is also important to notice that, as it happens for the purely metric case, any $f(R)$ theory with torsion (with or without spin) is equivalent to a corresponding torsional scalar--tensor theory with Brans--Dicke parameter $\omega_0=0$ and with specific potential for the scalar field directly induced by the function $f(R)$ \cite{CCSV1,CCSV2}.
\\
Moreover, recalling the expression \eqref{2.7} for the contorsion tensor and making use of equations \eqref{3.5b}, we obtain the following representations
\begin{equation}\label{3.5.1}
K_{ij}^{\;\;\;h}= \hat{K}_{ij}^{\;\;\;h} + \hat{S}_{ij}^{\;\;\;h}
\end{equation}
where one has
\begin{subequations}\label{3.5.2}
\begin{equation}\label{3.5.2a}
\hat{S}_{ij}^{\;\;\;h}:=\frac{1}{2f'}\/\left( - S_{ij}^{\;\;\;h} + S_{j\;\;\;i}^{\;\;h} - S^h_{\;\;ij}\right)
\end{equation}
\begin{equation}\label{3.5.2b}
\hat{K}_{ij}^{\;\;\;h} := -\hat{T}_j\delta^h_i + \hat{T}_pg^{ph}g_{ij}
\end{equation}
\begin{equation}\label{3.5.2c}
\hat{T}_j:=\frac{1}{2f'}\/\left( \de{f'}/de{x^j} + S^{\;\;\;q}_{jq} \right)
\end{equation}
\end{subequations}
for the decomposed tensors.
\\
Inserting equations \eqref{3.5.1} and \eqref{3.5.2} into equations \eqref{2.8}, we get the following useful decomposition of the Ricci tensor
\begin{equation}\label{3.5.3}
\begin{split}
R_{ij}=\tilde{R}_{ij} + \tilde{\nabla}_j\hat{K}_{hi}^{\;\;\;h} + \tilde{\nabla}_j\hat{S}_{hi}^{\;\;\;h} - 
\tilde{\nabla}_h\hat{K}_{ji}^{\;\;\;h} - \tilde{\nabla}_h\hat{S}_{ji}^{\;\;\;h} + \hat{K}_{ji}^{\;\;\;p}\hat{K}_{hp}^{\;\;\;h} + \hat{K}_{ji}^{\;\;\;p}\hat{S}_{hp}^{\;\;\;h} \\
+ \hat{S}_{ji}^{\;\;\;p}\hat{K}_{hp}^{\;\;\;h} + \hat{S}_{ji}^{\;\;\;p}\hat{S}_{hp}^{\;\;\;h} - \hat{K}_{hi}^{\;\;\;p}\hat{K}_{jp}^{\;\;\;h} - \hat{K}_{hi}^{\;\;\;p}\hat{S}_{jp}^{\;\;\;h} - \hat{S}_{hi}^{\;\;\;p}\hat{K}_{jp}^{\;\;\;h} - \hat{S}_{hi}^{\;\;\;p}\hat{S}_{jp}^{\;\;\;h}
\end{split}
\end{equation}
which we shall employ later.
\\
Now, we shall use the Bianchi identities \eqref{2.5} to work out the conservation laws for the above outlined theory. To see this point, let us consider the Bianchi identities in their fully contracted form as
\begin{subequations}
\label{3.6}
\begin{eqnarray}
\label{3.6a}
\nabla_{c}\left(R^{ci}-\frac{1}{2}g^{ci}R\right)= + \frac{1}{2}T_{cja}R^{cjai} + T^{ica}R_{ca}
\end{eqnarray}
\begin{eqnarray}
\label{3.6b}
\nabla_{c}\left(T^{ijc}+g^{ci}T^{j}-g^{cj}T^{i}\right) = - T^{ija}T_{a} - R^{ij} + R^{ji}
\end{eqnarray}
\end{subequations}
together with the field equations \eqref{3.5} written in the equivalent form
\begin{subequations}
\label{3.7}
\begin{equation}
\label{3.7a}
f'R^{ij}-\frac{1}{2}g^{ij}f=\Sigma^{ij}
\end{equation}
\begin{equation}
\label{3.7b}
f'\left(T^{tsa}-T^{t}g^{sa}+T^{s}g^{ta}\right)=\nabla^{t}f'g^{sa}-\nabla^{s}f'g^{ta}+S^{tsa}.
\end{equation}
\end{subequations}
The divergences of equations \eqref{3.7} read respectively as
\begin{subequations}
\label{3.8}
\begin{equation}
\label{3.8a}
\nabla_{i}f'R^{ij}+f'\nabla_{i}\left(R^{ij}-\frac{1}{2}g^{ij}R\right)=\nabla_{i}\Sigma^{ij}
\end{equation}
\begin{eqnarray}
\label{3.8b}
\nonumber
&\nabla_{a}f'\left(T^{tsa}-T^{t}g^{sa}+T^{s}g^{ta}\right)
+f'\nabla_{a}\left(T^{tsa}-T^{t}g^{sa}+T^{s}g^{ta}\right)=\\
&=[\nabla^{s},\nabla^{t}]f'+\nabla_{a}S^{tsa}
\end{eqnarray}
\end{subequations}
and by inserting the content of the fully contracted Bianchi identities \eqref{3.6} into equations \eqref{3.8}, we get
\begin{subequations}
\label{3.9}
\begin{equation}
\label{3.9a}
\nabla_{i}f'R^{ij}+\frac{1}{2}f'T_{cka}R^{ckaj}+f'T^{jca}R_{ca}=\nabla_{i}\Sigma^{ij}
\end{equation}
\begin{eqnarray}
\label{3.9b}
\nonumber
&\nabla_{a}f'\left(T^{tsa}-T^{t}g^{sa}+T^{s}g^{ta}\right)-f'T^{tsa}T_{a}
+f'\left(R^{st}-R^{ts}\right)=\\
&=[\nabla^{s},\nabla^{t}]f'+\nabla_{a}S^{tsa}
\end{eqnarray}
\end{subequations}
so that evaluating explicitly the term $[\nabla^{s},\nabla^{t}]f'$, the expressions \eqref{3.9} can be further simplified as
\begin{subequations}
\label{3.10}
\begin{equation}
\label{3.10a}
\nabla_{i}f'R^{ij}+\frac{1}{2}f'T_{cka}R^{ckaj}+f'T^{jca}R_{ca}=\nabla_{i}\Sigma^{ij}
\end{equation}
\begin{eqnarray}
\label{3.10b}
\nabla^{t}f'T^{s}-\nabla^{s}f'T^{t}-f'T^{tsa}T_{a}+f'\left(R^{st}-R^{ts}\right)=\nabla_{a}S^{tsa}.
\end{eqnarray}
\end{subequations}
At this point, the field equations \eqref{3.7} can again be used to calculate: the antisymmetric part of \eqref{3.7a}
\begin{subequations}
\label{3.11}
\begin{equation}
\label{3.11a}
f'\left(R^{ij}-R^{ji}\right)=\left(\Sigma^{ij}-\Sigma^{ji}\right)
\end{equation}
the contractions of \eqref{3.7a} with torsion vector and torsion 
\begin{equation}
\label{3.11b}
f'R^{ij}T_{i}-\frac{1}{2}fT^{j}=\Sigma^{ij}T_{i}
\end{equation}
\begin{equation}
\label{3.11c}
f'R^{ij}T_{bij}-\frac{1}{2}fT_{b}=\Sigma^{ij}T_{bij}
\end{equation}
the contractions of \eqref{3.7b} with curvature and torsion 
\begin{equation}
\label{3.11d}
\frac{1}{2}f'R_{tsai}T^{tsa}+T^{t}f'R_{ti}=-R_{ti}\nabla^{t}f'+\frac{1}{2}S^{tsa}R_{tsai}
\end{equation}
\begin{equation}
\label{3.11e}
f'T^{tsa}T_{a}=T^{s}\nabla^{t}f'-T^{t}\nabla^{s}f'+S^{tsa}T_{a}
\end{equation}
\end{subequations}
and by finally inserting the content of all equations \eqref{3.11} into equations \eqref{3.10}, we obtain the following conservation laws
\begin{subequations}
\label{3.12}
\begin{equation}
\label{3.12a}
\nabla_{a}\Sigma^{ai}+T_{a}\Sigma^{ai}-\Sigma_{ca}T^{ica}-\frac{1}{2}S_{sta}R^{stai}=0
\end{equation}
\begin{equation}
\label{3.12b}
\nabla_{h}S^{ijh}+T_{h}S^{ijh}+\Sigma^{ij}-\Sigma^{ji}=0
\end{equation}
\end{subequations}
under which the stress-energy and spin density tensors of the matter fields must undergo once the matter field equations are assigned.
\\
It is worth noticing that in the case of coupling with matter without spin, equations \eqref{3.12} reduces to the usual conservation laws holding in general relativity. Indeed, in metric-affine formulation with no spin equations \eqref{3.12b} are identities. At the same time, working out equation \eqref{3.12a} and taking equations \eqref{2.6} and \eqref{2.7} into account, we have
\begin{equation}\label{3.13}
0= \tilde{\nabla}_i\Sigma^{ij} - K_{ih}^{\;\;\;i}\Sigma^{hj} - K_{ih}^{\;\;\;j}\Sigma^{ih} +T_{h}\Sigma^{hj}-T^{j}_{\;\;ih}\Sigma^{ih} = \tilde{\nabla}_i\Sigma^{ij}.
\end{equation}
This last result has been already proved in a different way \cite{CV1,CV2,CV3}, exploiting the dynamical equivalence of metric-affine $f(R)$-gravity without spin with certain Brans-Dicke-like theories, for which the standard conservation laws given in equation \eqref{3.13} hold \cite{Koivisto}. The proof given above is more direct and geometrical.
\\
Next we shall investigate more in detail the coupling with Dirac fields.
\section{Coupling with Dirac matter fields and consistency}
Let us consider $f(R)$-gravity coupled with a Dirac field. The matter Lagrangian function is the Dirac one
\begin{equation}
\label{4.1}
{\cal L}_m=
\left[\frac{i}{2}\left(\bar\psi\gamma^iD_{i}\psi-D_{i}\bar\psi\gamma^{i}\psi\right)-m\bar\psi\psi\right]
\end{equation}
where $D_i\psi = \de\psi/de{x^i} + \omega_i^{\;\;\mu\nu}S_{\mu\nu}\psi\/$ and $D_i\bar\psi = \de{\bar\psi}/de{x^i} - \bar\psi\omega_i^{\;\;\mu\nu}S_{\mu\nu}\/$ are the covariant derivatives of the Dirac fields, $S_{\mu\nu}=\frac{1}{8}\left[\gamma_\mu,\gamma_\nu\right]\/$, $\gamma^i =\gamma^{\mu}e^i_\mu\/$ with $\gamma^\mu\/$ denoting Dirac matrices and where $m$ is the mass of the Dirac field. We use the notation for which 
\begin{equation}
\label{4.1.1}
\gamma^{\mu}\gamma^{\nu}\gamma^{\lambda}
=\gamma^{\mu}\eta^{\nu\lambda}-\gamma^{\nu}\eta^{\mu\lambda}+\gamma^{\lambda}\eta^{\mu\nu}
+i\epsilon^{\mu\nu\lambda\tau}\gamma_{5}\gamma_\tau
\end{equation}
which implicitly defines the $\gamma_5$ matrix and which entails the identities
\begin{subequations}
\label{4.1.2}
\begin{equation}
\label{4.1.2a}
[\gamma^{\mu},[\gamma^{\nu},\gamma^{\lambda}]]
=4\left(\gamma^{\lambda}\eta^{\mu\nu}-\gamma^{\nu}\eta^{\mu\lambda}\right)
\end{equation}
and
\begin{equation}
\label{4.1.2b}
\{\gamma^{\mu},[\gamma^{\nu},\gamma^{\lambda}]\}
=4i\epsilon^{\mu\nu\lambda\tau}\gamma_{5}\gamma_\tau
\end{equation}
\end{subequations}
we will employ in the following. 
\\
By varying \eqref{4.1} with respect to $\psi$, we obtain the matter field equations
\begin{equation}
\label{4.3}
i\gamma^{h}D_{h}\psi + \frac{i}{2}T_h\gamma^h\psi- m\psi=0
\end{equation}
which generalize the usual Dirac equations. Due to equations \eqref{4.3}, it is easily seen that the stress-energy and spin density tensors, derived as in section 3, are respectively given by
\begin{subequations}
\label{4.2}
\begin{equation}
\label{4.2a}
\Sigma_{ij}=\frac{i}{4}\left(\bar\psi\gamma_{i}D_{j}\psi-D_{j}\bar\psi\gamma_{i}\psi\right)
\end{equation}
\begin{equation}
\label{4.2b}
S_{ij}^{\;\;\;h}=\frac{i}{2}\bar\psi\left\{\gamma^{h},S_{ij}\right\}\psi
\equiv-\frac{1}{4}\eta^{\mu\sigma}\epsilon_{\sigma\nu\lambda\tau}
\left(\bar{\psi}\gamma_{5}\gamma^{\tau}\psi\right)e^{h}_{\mu}e^{\nu}_{i}e^{\lambda}_{j}
\end{equation}
\end{subequations}
where we have used \eqref{4.1.2b} to show the equivalence of the two forms of the spin density, which is then manifestly completely antisymmetric and thus irreducible. 
\\
Now, it is possible to calculate the divergences of the quantities \eqref{4.2} as
\begin{subequations}
\label{4.4}
\begin{equation}
\label{4.4a}
D_{i}\Sigma^{ij}
=\frac{i}{4}\left(D_{i}\bar\psi\gamma^{i}D^{j}\psi+\bar\psi\gamma^{i}D_{i}D^{j}\psi
-D_{i}D^{j}\bar\psi\gamma^{i}\psi-D^{j}\bar\psi\gamma^{i}D_{i}\psi\right)
\end{equation}
\begin{equation}
\label{4.4b}
D_{h}S^{ijh}
=\frac{i}{2}\left(D_{h}\bar\psi\gamma^{h}S^{ij}\psi+\bar\psi\gamma^{h}S^{ij}D_{h}\psi
+D_{h}\bar\psi S^{ij}\gamma^{h}\psi+\bar\psi S^{ij}\gamma^{h}D_{h}\psi\right)
\end{equation}
\end{subequations}
and by adding and removing the terms $\bar{\psi}\gamma^iD^jD_i\psi$ and $D^jD_i\bar{\psi}\gamma^i\psi$ in equations \eqref{4.4a} as well as the terms $\bar{\psi}S^{ij}\gamma^hD_h\psi$ and $D_h\bar{\psi}\gamma^hS^{ij}\psi$ in equations \eqref{4.4b}, we obtain the following expressions
\begin{subequations}
\label{4.5}
\begin{eqnarray}
\label{4.5a}
\nonumber
&D_{i}\Sigma^{ij}
=\frac{i}{4}(\bar\psi\gamma_{i}[D^{i},D^{j}]\psi-[D^{i},D^{j}]\bar\psi\gamma_{i}\psi+\\
&+D_{i}\bar\psi\gamma^{i}D^{j}\psi+\bar\psi\gamma_{i}D^{j}D^{i}\psi
-D^{j}D^{i}\bar\psi\gamma_{i}\psi-D^{j}\bar\psi\gamma^{i}D_{i}\psi)
\end{eqnarray}
\begin{equation}
\label{4.5b}
D_{h}S^{ijh}
=\frac{i}{2}(\bar\psi[\gamma^{h},S^{ij}]D_{h}\psi+D_{h}\bar\psi [S^{ij},\gamma^{h}]\psi
+2D_{h}\bar\psi\gamma^{h}S^{ij}\psi+2\bar\psi S^{ij}\gamma^{h}D_{h}\psi).
\end{equation}
\end{subequations}
By employing the matter field equations \eqref{4.3}, we get
\begin{subequations}
\label{4.6}
\begin{equation}
\label{4.6a}
D_{i}\Sigma^{ij}
=\frac{i}{4}(\bar\psi\gamma_{i}[D^{i},D^{j}]\psi-[D^{i},D^{j}]\bar\psi\gamma_{i}\psi + D^j\bar{\psi}\gamma^h\psi\/T_h - T_h\bar{\psi}\gamma^hD^j\psi)
\end{equation}
\begin{equation}
\label{4.6b}
D_{h}S^{ijh}
=\frac{i}{2}(\bar\psi[\gamma^{h},S^{ij}]D_{h}\psi+D_{h}\bar\psi [S^{ij},\gamma^{h}]\psi - T_h\bar{\psi}\{\gamma^h,S^{ij}\}\psi).
\end{equation}
\end{subequations}
Finally, by making use of the commutator of covariant derivatives on spinors and the algebraic identities given by \eqref{4.1.2a}, we have
\begin{subequations}
\label{4.7}
\begin{equation}
\label{4.7a}
D_{i}\Sigma^{ij}=- T_i\Sigma^{ij} -T^{ijk}\Sigma_{ik}+\frac{1}{2}S_{abi}R^{abij}
\end{equation}
\begin{equation}
\label{4.7b}
D_{h}S^{ijh}=- T_hS^{ijh} + (\Sigma^{ji}-\Sigma^{ij})
\end{equation}
\end{subequations}
proving the complete consistency between the matter field equations \eqref{4.3} and the corresponding conservation laws \eqref{3.12}.
\\
We now discuss more in detail the Einstein-like equations of the theory, separating the Levi--Civita contributions from the torsional ones. To this end, due to the total antisymmetry of the spin density, we note that the Ricci tensor and the curvature scalar can be expressed as
\begin{subequations}\label{4.8}
\begin{equation}\label{4.8a}
R_{ij} = \tilde{R}_{ij} - 2\tilde{\nabla}_{j}\hat{T}_i - \tilde{\nabla}_h\hat{T}^hg_{ij} + 2\hat{T}_i\hat{T}_j - 2\hat{T}_h\hat{T}^hg_{ij} - \tilde{\nabla}_h\hat{S}_{ji}^{\;\;\;h} - \hat{S}_{hi}^{\;\;\;p}\hat{S}_{jp}^{\;\;\;h}
\end{equation}
and also
\begin{equation}\label{4.8b}
R = \tilde{R} - 6\tilde{\nabla}_{i}\hat{T}^i - 6\hat{T}_i\hat{T}^i - \hat{S}_{hi}^{\;\;\;p}\hat{S}_{\;\;p}^{i\;\;\;h}
\end{equation}
\end{subequations}
where one has $\hat{T}_i =\frac{1}{2f'}\de{f'}/de{x^i}\/$ e $\hat{S}_{ij}^{\;\;\;h}=-\frac{1}{2f'}S_{ij}^{\;\;\;h}\/$ (compare with equations \eqref{3.5.2} and \eqref{3.5.3}). Taking the decompositions \eqref{4.8} into account, from equations \eqref{3.5a} we derive the Einstein-like equations
\begin{equation}\label{4.9}
\begin{split}
\tilde{R}_{ij} -\frac{1}{2}\tilde{R}g_{ij}= \frac{1}{\varphi}\Sigma_{ij} + \frac{1}{\varphi^2}\left( - \frac{3}{2}\de\varphi/de{x^i}\de\varphi/de{x^j} + \varphi\tilde{\nabla}_{j}\de\varphi/de{x^i} + \frac{3}{4}\de\varphi/de{x^h}\de\varphi/de{x^k}g^{hk}g_{ij} \right. \\
\left. - \varphi\tilde{\nabla}^h\de\varphi/de{x^h}g_{ij} - V\/(\varphi)g_{ij} \right) + \tilde{\nabla}_h\hat{S}_{ji}^{\;\;\;h} + \hat{S}_{hi}^{\;\;\;p}\hat{S}_{jp}^{\;\;\;h} - \frac{1}{2}\hat{S}_{hq}^{\;\;\;p}\hat{S}_{\;\;p}^{q\;\;\;h}g_{ij}
\end{split}
\end{equation}
where the scalar field
\begin{equation}\label{4.10}
\varphi := f'\/(F\/(\Sigma))
\end{equation}
and the effective potential
\begin{equation}\label{4.11}
V\/(\varphi):= \frac{1}{4}\left[ \varphi F^{-1}\/((f')^{-1}\/(\varphi)) + \varphi^2\/(f')^{-1}\/(\varphi)\right]
\end{equation}
have been introduced \cite{CCSV1,CCSV2}. Splitting the symmetric part of \eqref{4.9} from the antisymmetric one, we get the following two sets of equations 
\begin{subequations}\label{4.12}
\begin{equation}\label{4.12a}
\begin{split}
\tilde{R}_{ij} -\frac{1}{2}\tilde{R}g_{ij}= \frac{1}{\varphi}\Sigma_{(ij)} + \frac{1}{\varphi^2}\left( - \frac{3}{2}\de\varphi/de{x^i}\de\varphi/de{x^j} + \varphi\tilde{\nabla}_{j}\de\varphi/de{x^i} + \frac{3}{4}\de\varphi/de{x^h}\de\varphi/de{x^k}g^{hk}g_{ij} \right. \\
\left. - \varphi\tilde{\nabla}^h\de\varphi/de{x^h}g_{ij} - V\/(\varphi)g_{ij} \right) + \hat{S}_{hi}^{\;\;\;p}\hat{S}_{jp}^{\;\;\;h} - \frac{1}{2}\hat{S}_{hq}^{\;\;\;p}\hat{S}_{\;\;p}^{q\;\;\;h}g_{ij}
\end{split}
\end{equation}
\begin{equation}\label{4.12b}
\frac{1}{\varphi}\Sigma_{[ij]} + \tilde{\nabla}_h\hat{S}_{ji}^{\;\;\;h} = 0
\end{equation}
\end{subequations}
and it is a straightforward matter (left to the reader) to verify that equations \eqref{4.12b} are equivalent to equations \eqref{4.7b}. In other words, the matter field equations \eqref{4.3} imply equations \eqref{4.12b} and the significant part of the Einstein-like equations reduces to the equations \eqref{4.12a} only. The latter can be worked out by giving an explicit representation of the covariant derivative of spinor fields. To this end, we denote by $\tilde{D}$ the spinorial covariant derivative induced by the Levi--Civita connection. Then, we have 
\begin{subequations}\label{4.13}
\begin{equation}\label{4.13a}
D_i\psi = \tilde{D}_i\psi - \frac{1}{4}\left(\hat{K}_{ijh} + \hat{S}_{ijh}\right)\gamma^h\gamma^j\psi
\end{equation}
\begin{equation}\label{4.13b}
D_i\bar\psi = \tilde{D}_i\bar\psi + \frac{1}{4}\bar{\psi}\left(\hat{K}_{ijh} + \hat{S}_{ijh}\right)\gamma^h\gamma^j
\end{equation}
\end{subequations}
which can be used together with of \eqref{2.6}, \eqref{3.5.1}, \eqref{3.5.2} and \eqref{4.2b}, to obtain the following identities
\begin{equation}\label{4.14}
\begin{split}
\bar\psi\gamma_iD_j\psi - \left(D_j\bar\psi\right)\gamma_i\psi = \bar\psi\gamma_i\tilde{D}_j\psi - (\tilde{D}_j\bar\psi)\gamma_i\psi -\frac{i}{8\varphi}(\bar{\psi}\gamma_5\gamma^\tau\psi)(\bar{\psi}\gamma_5\gamma_\tau\psi)g_{ij} \\
+ \frac{i}{8\varphi}(\bar{\psi}\gamma_5\gamma_i\psi)(\bar{\psi}\gamma_5\gamma_j\psi) + \frac{i}{2\varphi}\partial_{h}\varphi\epsilon^\nu_{\;\sigma\alpha\tau}e^h_\nu\/e^\sigma_i\/e^\alpha_j(\bar{\psi}\gamma_5\gamma^\tau\psi)
\end{split}
\end{equation}
\begin{equation}\label{4.15}
\hat{S}_{hi}^{\;\;\;p}\hat{S}_{jp}^{\;\;\;h} = -\frac{1}{32(\varphi)^2}
(\bar{\psi}\gamma_5\gamma^\tau\psi)(\bar{\psi}\gamma_5\gamma_\tau\psi)g_{ij}
+ \frac{1}{32(\varphi)^2}
(\bar{\psi}\gamma_5\gamma_i\psi)(\bar{\psi}\gamma_5\gamma_j\psi)
\end{equation}
\begin{equation}\label{4.16}
- \frac{1}{2}\hat{S}_{hq}^{\;\;\;p}\hat{S}_{\;\;p}^{q\;\;\;h}g_{ij} = \frac{3}{64(\varphi)^2}(\bar{\psi}\gamma_5\gamma^\tau\psi)(\bar{\psi}\gamma_5\gamma_\tau\psi)g_{ij}
\end{equation}
and finally by inserting \eqref{4.14}, \eqref{4.15} e \eqref{4.16} into \eqref{4.12} we get the expression for the symmetrized Einstein-like equations
\begin{equation}
\label{4.18}
\begin{split}
\tilde{R}_{ij} -\frac{1}{2}\tilde{R}g_{ij}= \frac{1}{\varphi}\frac{i}{4}\/\left[ \bar\psi\gamma_{(i}\tilde{D}_{j)}\psi - \left(\tilde{D}_{(j}\bar\psi\right)\gamma_{i)}\psi \right]
+ \frac{1}{\varphi^2}\left( - \frac{3}{2}\de\varphi/de{x^i}\de\varphi/de{x^j} + \varphi\tilde{\nabla}_{j}\de\varphi/de{x^i} + \right. \\
\left. \frac{3}{4}\de\varphi/de{x^h}\de\varphi/de{x^k}g^{hk}g_{ij}
- \varphi\tilde{\nabla}^h\de\varphi/de{x^h}g_{ij} - V\/(\varphi)g_{ij} \right) + \frac{3}{64\varphi^2}(\bar{\psi}\gamma_5\gamma^\tau\psi)(\bar{\psi}\gamma_5\gamma_\tau\psi)g_{ij}.
\end{split}
\end{equation}
Next, we have that Dirac equations can be decomposed in torsionless and torsional terms; indeed, by using the identity $\left[(\bar{\psi}\gamma_5\gamma^\tau\psi)\gamma_5+(\bar{\psi}\gamma^\tau\psi)\mathbb{I}\right] \gamma_\tau\psi\equiv0$ together with the Fierz rearrangement, we get the following equivalent form for Dirac equations 
\begin{equation}
\label{4.20}
i\gamma^{h}\tilde{D}_{h}\psi
-\frac{1}{\varphi}\frac{3}{16}\left[(\bar{\psi}\psi)
+i(i\bar{\psi}\gamma_5\psi)\gamma_5\right]\psi-m\psi=0
\end{equation}
where the coupling between the spin density and the completely antisymmetric irreducible part of torsion generates spinorial self-interactions whose potentials can be recognized to be of the Nambu-Jona--Lasinio type \cite{n-j--l/1,n-j--l/2}.
\\
As a final step, by making use of equations \eqref{4.20} the symmetrized Einstein-like equations can be written in the equivalent form
\begin{equation}
\label{4.19}
\begin{split}
\tilde{R}_{ij}=\frac{1}{\varphi}\frac{i}{4}\/\left[ \bar\psi\gamma_{(i}\tilde{D}_{j)}\psi - \left(\tilde{D}_{(j}\bar\psi\right)\gamma_{i)}\psi \right] - \frac{1}{\varphi}\frac{m}{4}(\bar{\psi}\psi)g_{ij} \\
+ \frac{1}{\varphi^2}\left( - \frac{3}{2}\de\varphi/de{x^i}\de\varphi/de{x^j} + \varphi\tilde{\nabla}_{j}\de\varphi/de{x^i} + \frac{1}{2}\varphi\tilde{\nabla}^h\de\varphi/de{x^h}g_{ij} + V\/(\varphi)g_{ij} \right) 
\end{split}
\end{equation}
in which all torsional contributions have disappeared. We observe that the non-linearity of the $f(R)$ function gives rise to the appearance of the scalar field we indicated with $\varphi=\varphi(\bar{\psi}\psi)$ in the field equations \eqref{4.20} and \eqref{4.19}. The role of the scalar field $\varphi$ is two-fold: on the one hand, it provides additional terms in the effective stress-energy tensor in the Einstein-like field equations \eqref{4.19}; on the other hand, it is a scale factor modifying the normalization of the Dirac field contributions in both equations \eqref{4.20} and \eqref{4.19}. Furthermore, we also remark that the scalar $\varphi$ may assume both positive and negative values, but it cannot switch between them due to the fact that it cannot vanish (see for instance equations \eqref{4.20} and \eqref{4.19}). This has again two implications: it may determine both repulsive and attractive potentials by controlling the sign of the self-interacting term in equations \eqref{4.20}; it controls the sign of the stress-energy tensor of the Dirac field in equations \eqref{4.19}. This means that both particles and antiparticles may have positive energies in classical gravity.
\\
To explain this point, consider the circumstance in which $f(R)=R-\varepsilon R^{2}$, where $\varepsilon$ is a fixed positive parameter. In this situation, we have $\varphi=1+\varepsilon \ m (\bar{\psi}\psi)$, which clearly shows that the sign of $\varphi$ depends on the scalar $\bar{\psi}\psi$ built from the spinor field: for particles, we have $\bar{\psi}\psi\geqslant0$ so that $\varphi\geqslant0$ and the energy is positive; for antiparticles, we have $\bar{\psi}\psi\leqslant0$ so that for $\left|\bar{\psi}\psi\right|\leqslant\frac{1}{\varepsilon m}$ it is $\varphi\geqslant0$ and the energy is negative, but for $\left|\bar{\psi}\psi\right|\geqslant\frac{1}{\varepsilon m}$ it is $\varphi\leqslant0$ and the energy is positive. The consequence of this fact is that, for extremely high-density states, particles and antiparticles have a similar behaviour both with positive energies, but as the density decreases, their behaviour differentiates, and while particles maintain positive energy, we have that antiparticles would turn out to have negative energy. This mechanism accounts for a spontaneous symmetry breaking that takes place in classical physics. For example, in the instance of cosmological applications, we have that although in the beginning of the universe the particle-antiparticle duality and the positivity of their energies might have been an obvious fact, with the following expansion particles and antiparticles lose their symmetry, and whereas the former are still allowed, the latter are forced to disappear: that the particle-antiparticle duality be broken is not a new feature for torsional gravitational theories \cite{a-s,a-g-s1,a-g-s2}; on the other hand torsional non-linear $f(R)$-gravity appear to amplify such a mechanism, as we have just discussed here, with no need of any quantization whatsoever.
\\
As a conclusive remark, we stress a consequence of having associated particles and antiparticles respectively with positive and negative values of the scalar field $\varphi$: recalling that the scalar $\varphi$ is allowed to have both positive and negative values but it cannot switch sign, then the model is allowed to describe both particles and antiparticles but it cannot provide a mechanism for transmuting them into each other by means of gravitational effects alone. As the theory describes two sub-theories, one for positive and one for negative $\varphi$, without the possibility to turn one into the other, then the Dirac matter field gives two kind of solutions, particles and antiparticles, without the possibility to change into one another just in terms of gravitational physics, compatibly with all observations made so far.
\section{Conclusions}
In this paper, we have considered $f(R)$-theories of gravitation with Ricci scalar written in terms of connections having both metric and torsional degrees of freedom. We have studied them coupled to Dirac matter fields: we have seen that for general $f(R)$ functions with $R$ depending on metric and torsion, the corresponding field equations transform the Bianchi geometrical identities into conservation laws; these conservation laws are satisfied for the stress-energy and spin density tensors of the Dirac field, once Dirac matter field equations are used.
\\
After the set of field equations has been checked for consistency, we have decomposed all equations in terms of torsionless and torsional contributions. We have seen that within the fermionic field equations the self-interacting term is given by a potential of the Nambu-Jona--Lasinio form scaled in terms of a scalar field $\varphi$; the Einstein-like field equations are such that the fermionic contributions are also scaled in terms of the same scalar field $\varphi$, and furthermore an additional stress-energy tensor for the scalar field $\varphi$ itself also appears. In this situation, we have discussed peculiar circumstances given by specific examples of $\varphi$ for which both particles and antiparticles have positive energies, displaying a duality in the beginning of the universe that is eventually lost in further stages of the cosmic evolution. It is remarkable that the positivity of the energy for both particles and antiparticles is ensured and the particle-antiparticle duality is gravitationally spontaneously broken with a classic approach. The fact that no quantization, let alone quantization with anticommutators, is considered at all is important for the foundations of quantum field theory in view of the problems quantum field theory is known to have. In fact, the concern about divergences that are introduced as a byproduct of the quantization of fields can be circumvented by avoiding the quantization of fields in the first place, as long as the problems of the positivity of energies and the particle-antiparticle duality solved by quantizing matter fields find solution in an alternative way. One of these alternative is provided here.

\end{document}